\documentclass[twocolumn,showpacs,preprintnumbers,amsmath,amssymb]{revtex4}
\pdfoutput=1

\usepackage{xspace}
\usepackage{graphicx}

\def\beq{\begin{equation}}
\def\eeq{\end{equation}}

\def\bea{\begin{eqnarray}}
\def\eea{\end{eqnarray}}

\newcommand{\roughly}[1]%
    {{\mathrel{\raise.3ex\hbox{$#1$\kern-.75em\lower1ex\hbox{$\sim$}}}}}
\newcommand{\lsim}{\mathrel{\roughly<}}

\newcommand{\scr}[1]{\ensuremath{\mathcal{#1}}}

\newcommand{\al}{\ensuremath{\alpha}}

\newcommand{\ga}{\ensuremath{\gamma}}
\newcommand{\Ga}{\ensuremath{\Gamma}}

\newcommand{\si}{\ensuremath{\sigma}}

\newcommand{\Om}{\ensuremath{\Omega}}

\newcommand{\GeV}{\ensuremath{\mathrm{~GeV}}}
\newcommand{\TeV}{\ensuremath{\mathrm{~TeV}}}

\newcommand{\Eq}[1]{Eq.~(\ref{#1})}

\newcommand{\Ref}[1]{Ref.~\cite{#1}}

\newcommand{\met}{\mbox{${\rm \not\! E}_{\rm T}$}}

\begin{document}


\title{Displaced Dark Matter at Colliders}

\author{Spencer Chang}
\email{schang@physics.ucdavis.edu}
\author{Markus A. Luty}%
\email{luty@physics.ucdavis.edu}
\affiliation{%
Physics Department,~University of California Davis\\
Davis,~California 95616}%


\begin{abstract}
Models in which the dark matter is very weakly coupled
to the observable sector may explain the observed dark
matter density, either as a ``superWIMP''
or as ``asymmetric dark matter.''
Both types of models predict displaced
vertices at colliders, with a rich variety of possible
phenomenology.
We classify the cases in which the decays can naturally occur inside
particle detectors at the LHC,
with particular focus on the nontrivial scenarios where the
decaying particle is invisible.  
Identification of the position and timing of these invisible displaced
vertices significantly improves the prospects of reconstructing
the new physics in models such as supersymmetry.
In many cases, reconstruction of the visible products
of the displaced decay can determine the dark matter
mass, allowing the dark matter density to be predicted from
collider data.
\end{abstract}


\maketitle

\section{\label{sec:intro}Introduction}
The nature of the dark matter that accounts for about
one quarter of the mass of the universe is one of the
most important open questions in particle physics and cosmology.
One simple possibility is that the dark matter is a massive particle $X$
whose relic density today results from the freeze-out of
the interactions $\bar{X}X \leftrightarrow \mbox{SM}$,
where SM are standard model particles.
In this case, the relic density is \cite{Kolb:1990vq}
\beq
\label{WIMPmiracle}
\Om_X \sim \frac{\mbox{pb}}{\si(\bar{X}X \to \mbox{SM})}.
\eeq
That is, the correct relic abundance $\Om_X \sim 1$ is obtained
for an annihilation cross section typical for a perturbative
interaction of a weak-scale particle.
Such particles are called weakly interacting massive particles
(WIMPs), and \Eq{WIMPmiracle} is sometimes called the ``WIMP miracle.''
It suggests that dark matter may be connected with the weak
scale, and hence may be observable at the LHC.

Another hint for the existence of WIMPs comes from
attempts to understand particle physics at the weak scale.
Particle physics models that address the stability of the weak
scale against quantum corrections generally have
``partners'' of standard model particles with masses of order
$100\GeV$--1 TeV whose role
is to cancel the UV sensitivity of the electroweak order parameter.
The most well-studied example is that of supersymmetry (SUSY),
with a SUSY partner for every standard model
particle, but ``little Higgs'' and models with extra dimensions
also fall into this class.
Furthermore, these models naturally satisfy precision electroweak
constraints if the partner particles cannot be singly produced,
implying an approximate ``partner parity'' under which the 
partners are odd and standard model particles are even.
If this parity symmetry is exact, the lightest odd particle
is absolutely stable, and generically
has the mass and interactions to be
WIMP dark matter.

If this picture is correct, the
prospects for obtaining experimental evidence for it are excellent.
Dark matter direct and indirect detection experiments are reaching
levels of sensitivity where a signal is expected.
Also, the CERN Large Hadron Collider (LHC) will explore
the weak scale with $pp$ collisions with energies up to $14\TeV$,
leading to the exciting possibility that
dark matter can be produced and studied in the laboratory.
However, the connection between dark matter production
at colliders and the cosmological dark matter density
is very indirect.
The production of WIMPs at colliders occurs via pair
production of the standard model partners with the strongest
couplings to protons (typically colored partners),
followed by cascade decays to standard model particles
and dark matter particles (see Fig.~1a).
This is not directly related to the $\bar{X} X$
annihilation process that
determines the cosmological relic density,
and knowledge of the underlying model is required to relate them.
The program of relating collider data to cosmological dark matter is
extremely ambitious, and may require
colliders beyond the LHC
({\it e.g.\/}\ a high-energy $e^+ e^-$ collider) \cite{Baltz:2006fm}.

\begin{figure}[t]
\includegraphics{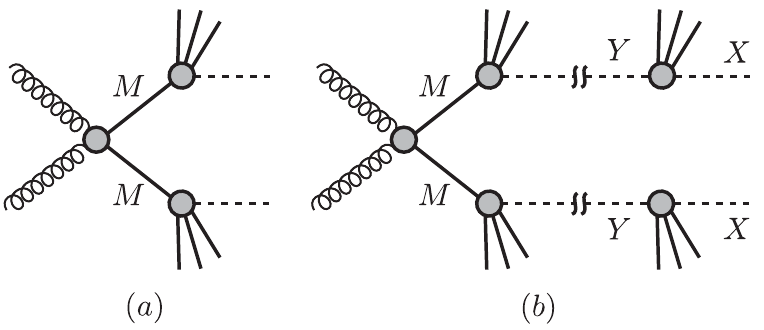}
\caption{\label{fig:collider} 
Collider production of dark matter (a) in
conventional WIMP dark matter models, (b) in 
superWIMP/ADM models.}
\end{figure}

However, there are other models of dark matter
in which there is an explanation of the dark matter
as compelling as \Eq{WIMPmiracle}, and
also fit well with particle physics at the weak scale,
but where the program of relating collider and cosmological
data is at first sight even more difficult, perhaps even impossible.
In this Letter we
show that in many cases these involve displaced
vertices visible in particle detectors, and these 
substantially improve the prospects for connecting
collider and cosmological data.
We will discuss these ideas in the context of
SUSY, specifically the minimal supersymmetric standard model
(MSSM), but it is straightforward to extend
them to other particle theory frameworks. 
We consider two possibilities in particular.

{\bf SuperWIMP models}~\cite{Feng:2003xh}:
In these models, the lightest supersymmetric particle
in the visible sector is the
next-to-lightest SUSY partner (NLSP) $Y$,
and decays to a SUSY particle $X$ in the dark sector.
It is assumed that $X$ is sufficiently weakly coupled 
that it is never in thermal equilibrium, in which case
its relic density is set by the decay of $Y$ particles
after they freeze out.
We then obtain
\beq
\Om_X \sim \frac{m_X}{m_Y}
\frac{\mbox{pb}}{\si(\bar{Y}Y \rightarrow \mbox{SM})}.
\eeq
This preserves the motivation of \Eq{WIMPmiracle}
as long as $m_Y \sim m_X$.
This is natural if the masses of both $X$ and $Y$ arise
from breaking SUSY or electroweak symmetry.

{\bf Asymmetric dark matter (ADM) models}~\cite{ADM}:
In these models the 
dark matter density arises from a
relic particle-antiparticle asymmetry in the dark
matter particles $X$.
The visible sector has a baryon and lepton asymmetry
in the early universe that explains the fact that the universe
is composed of matter (and not antimatter) today.
These asymmetries may be related
\cite{Kaplan:1991ah,Dodelson:1991iv},
giving a common origin of the baryon and dark matter
relic densities.
In ADM models the lepton and baryon
asymmetry is transferred to $X$ via interactions that are in
equilibrium in the early universe but drop out of equilibrium 
at a temperature $T \gtrsim m_X$.
The $X$ density is then related to the baryon
density by
\beq
\frac{\Om_X}{\Om_B} \sim 
\frac{m_X}{m_B},
\eeq
where
$m_B \sim 1\GeV$ is the baryon mass.
This prediction only depends on the $X$ mass,
giving the correct dark matter abundance for $m_X \sim 10\GeV$,
close to the weak scale.

Remarkably, both classes of models predict displaced vertices
from NLSP decays (see Fig.~1b).
Dark matter therefore gives another motivation for displaced decay
searches in addition to those in the literature
({\it e.g\/}.~Ref.~\cite{Dimopoulos:1996vz}).  
In superWIMP models the long-lived NLSP arises because
$Y$ transfers its relic abundance to $X$ by decaying
after it freezes out at a temperature $T_Y \sim m_Y/25$.
This implies
\beq
\label{superWIMPdecaybound}
\Ga(Y \to X \cdots ) \lsim H(T_Y) \sim \frac{c}{10~\mbox{m}}\left(\frac{m_Y}{100 \GeV}\right)^2.
\eeq
This corresponds to a typical displaced vertex of the order of the
size of an LHC detector, another interesting example of a
``cosmic coincidence.''
The literature on superWIMPs has focused on decays via gravitational
strength couplings with much longer decay lengths, and the
possibility of decays inside the detector was not emphasized.

In ADM models, 
the light dark matter particle is generally the LSP
and the operator that transfers the particle-antiparticle
asymmetry to the dark sector allows the NLSP to decay.
This decay must be out of equilibrium at temperatures
$T \sim m_X \sim 10\GeV$,
otherwise the asymmetry in the dark sector is diluted.
This again gives a displaced vertex of order $10$~m
for $m_Y \sim 100\GeV$.

It is of course possible that $c\tau$ of the NLSP
is much longer than the detector radius $R$.
However even in this case the NLSP
will decay inside the detector
with probability $R/(\ga \beta c \tau)$.
The parameter space for models with observable
displaced vertices therefore extends to models with decay
lengths much longer than 10~m
(see {\it e.g.\/}~\Ref{Ishiwata:2008tp}).
We will show that the observation of even a few
displaced decays substantially improves the prospects
of reconstructing the events and measuring the parameters
relevant for cosmology.
If the NLSP is electrically charged, it can be tracked
to reconstruct the event fully, and it may be trapped and
its late decays studied 
\cite{Hamaguchi:2004df,Feng:2004yi}.
Our results are therefore of particular interest for the case
of uncharged NLSPs.


\section{Displaced Decay Phenomenology}

We begin by classifying the models in which the decay
lengths can saturate the bound \Eq{superWIMPdecaybound}.
In superWIMP models $X$ must be out of equilibrium at
temperatures above the $Y$ freeze-out temperature.
This happens naturally for renormalizable couplings,
which decouple at high temperatures.
Non-renormalizable couplings would have to be out of
equilibrium at the reheat temperature, effectively the highest
temperature reached by the universe, resulting in much
longer minimum decay lengths.
In the MSSM, the only possible renormalizable couplings
to a neutral superWIMP $X$ are the superpotential couplings
\beq
X H_u H_d
\quad\mbox{and}\quad
X L H_u.
\eeq
For the first operator $X$ is $R$ parity even
and the dark matter is the fermion component of $X$;
for the second $X$ is $R$ parity odd and the
dark matter is the scalar component.
Depending on supersymmetry breaking, it can be natural for the scalar component to be lighter
than the fermion component.  Assuming that the NLSP decays after its freeze out, 
scattering processes such as $\tilde{H} \tilde{H} \to X \tilde{W}$ were never in equilibrium, unlike nonrenormalizable couplings which at high enough temperatures eventually equilibrate.      

For ADM models, 
we require that the interaction that transfers the
asymmetry be \emph{in} equilibrium at sufficiently
high temperatures, so this is required to be a
non-renormalizable interaction.
This interaction generally allows the NLSP to decay
to one or more $X$ particles, as we will see below.
The standard model part of the coupling must carry
nonzero $B-L$.
The simplest possible operators are those of
dimension 5, all with $B-L = \pm 1$.
These are superpotential terms of the form
\beq
\label{WADMops}
X L L e^c,
\ \ 
X Q L d^c,
\ \ 
X u^c d^c d^c,
\ \ 
X^2 L H_u,
\eeq
and K\"ahler terms of the form
\beq
X L H_d^\dagger, 
\ \ 
X^\dagger L H_d^\dagger,
\ \ 
X^\dagger L H_u.
\eeq
\Ref{ADM} only considered interactions proportional to $X^2$,
but this is not required by the ADM mechanism.
With the exception of the last operator in \Eq{WADMops},
$X$ is $R$ parity odd and the dark matter is the $X$ scalar.
For the last operator in \Eq{WADMops} there is a
$Z_4$ $R$ symmetry that guarantees stability of $X$,
and the dark matter can be either the scalar or fermion component of $X$;
we consider the case where it is the scalar so that the NLSP
decays directly into the dark matter.
To obtain the correct dark matter density, the 
$X$ mass must be $\sim 6\GeV$ for all of these operators except for
the $X^2 L H_u$ model, which has an $X$ mass of $12\GeV$ \cite{ADM}.
More generally, 
one can show that for a coupling of the form $X^n \scr{O}$
one has
$m_X \simeq (6~\mbox{GeV}) n / |B - L|_{\cal O}$.
This assumes that the operator transferring the asymmetry freezes out
at a temperature {\em above} $m_X$.
If it freezes out at a lower temperature,
the predicted mass will increase and the NLSP decay length will get shorter.


Because the minimum decay length for the NLSP is already close
to the size of a detector, we focus on cases where the
NLSP field appears in the operator that couples the sectors.
Then there is no further suppression of the NLSP
decay, so these are the models most likely to have observable
displaced vertices.
Given this, the possible observable NLSP decays resulting from either
superWIMP and ADM models is
listed in Table~1.   
We have allowed the possibility of chargino NLSPs,
which can occur in some situations \cite{Kribs:2008hq}.
We have not listed completely invisible decays,
{\it e.g\/}.~$\chi^0 \to \nu X$ for the operator
$X L H_u$.
If we allow operators with dimension greater than 5, there are 
additional possibilities, such as $\tilde{\nu} \to X \ga\ga$
from the operator $X L H_u W^\al W_\al$, where
$W_\al$ is a standard model gauge field strength superfield.

\begin{table}[t] 
\begin{center}
\begin{tabular}{|c|c|c|}
\hline
Coupling & NLSP Decay & NLSP Signal \\
\hline
\hline
& $\chi^0 \to X + (h^0,Z)$ & $\text{vertex} \rightarrow (h^0,Z) +\met $\\[-.22cm]
$X H_u H_d$ & &\\[-.22cm]                     
& $\chi^\pm \to X + (H^\pm,W^\pm)$ & $\text{track} \rightarrow (H^\pm,W^\pm) +\met $\\[.1cm]
\hline
$X L H_u$& $\chi^\pm \to X + \ell^\pm $ & $\text{track} \rightarrow \ell^\pm +\met $\\
 $X L H_d^\dagger$ &  $\tilde{\nu} \to X + (h^0,Z)$ & $\text{vertex} \rightarrow (h^0,Z) +\met $\\
 $X^{\dagger} L H_u$ &  $\tilde{\ell}^\pm \to X + (H^\pm,W^\pm)$ & $\text{track} \rightarrow (H^\pm,W^\pm) +\met $\\[.1cm]
\hline
 & $\chi^\pm \to 2 X + \ell^\pm $ & $\text{track} \rightarrow \ell^\pm +\met $\\[-.22cm]
$X^2 L H_u$ & & \\[-.22cm]
 &  $\tilde{\ell}^\pm \to 2 X + (H^\pm,W^\pm)$ & $\text{track} \rightarrow (H^\pm,W^\pm) +\met $\\[.1cm]
\hline
&  $\tilde{\ell}^\pm \to X + \ell'^\pm$ & $\text{track} \rightarrow \ell'^\pm +\met $\\[-.22cm]
$X LL e^c$ & &\\[-.22cm]                     
& $\tilde{\nu} \to X + \ell'+\bar{\ell}$ & $\text{vertex} \rightarrow \ell'+\bar{\ell} +\met $\\[.1cm]
\hline
&$\tilde{\ell}^\pm \to X + u + \bar{d}  $ & $\text{track} \rightarrow 2~\mbox{jets} +\met $\\
& $\tilde{\nu} \to X + d +\bar{d}$ & $\text{vertex} \rightarrow  2~\mbox{jets} +\met $\\[-.22cm]
$X QLd^c$ & &\\[-.22cm]                     
& $\tilde{u} \to X + \ell^+ + d$ & $R\text{-hadron} \rightarrow \ell^\pm + \mbox{jet} +\met $\\
& $\tilde{d} \to X + \nu + d$ & $R\text{-hadron} \rightarrow \mbox{jet} +\met $\\[.1cm]                         
\hline
$X u^c d^c d^c$   & $\tilde{u} \to X + \bar{d} +\bar{d'}$ & $R\text{-hadron} \rightarrow 2~\mbox{jets} +\met $\\[.1cm]
\hline
\end{tabular}
\caption{NLSP decays and signals for various operators
coupling the dark matter to the MSSM.
The couplings $X L H_d^\dagger$ and $X^{\dagger} L H_u$
are K\"ahler couplings, the rest are
superpotential couplings.
For the coupling $X^{\dagger} L H_u$,
only the scalar NLSP decays can occur.   Flavor indices have been suppressed and depend on the flavor structure of the operator that $X$ couples to.
\label{table:decays}}
\end{center}
\end{table}


\section{Reconstruction}

Reconstructing decays with displaced vertices
is challenging at the LHC because the experiments
are designed to detect particles
coming from the interaction region.
The visible decay products from a highly displaced vertex
can travel in any direction through the detector,
and require special techniques.
These events should trigger on objects produced
in the supersymmetric prompt cascade, so at least
we can focus on offline reconstruction of these decays.

{\bf Prompt reconstruction:}
If the NLSP is charged or an $R$-hadron, measurements of
ionization energy loss and track curvature can determine
their 4-momenta (see \cite{Fairbairn:2006gg} for a review),
in principle allowing complete reconstruction of
the prompt part of the event.  
We therefore focus on the case of neutral NLSPs.
We can gain important information about the prompt part of
the event by determining the position and timing of the
displaced vertex, without fully reconstructing
the visible decay products of the vertex.

The position of the primary interaction vertex
can be determined using conventional methods
from the large number of 
hard visible particles expected from the
cascade decays (see Fig.~1b).
The position of the displaced vertex can in principle
be determined in many parts of the detector.
It is important to make use of the outer parts of the detector
since the probability to decay in a region of the detector is
essentially proportional to its radial size.
In the tracker, the displaced vertex can be located using
conventional tracking.
In outer parts of the detector, the displaced decay would give
activity without inner tracks pointing to it.
There may be distinguishing geometric features, such as
tracks that do not point radially, and studies would be needed
to determine whether the vertex can be identified in
these situations.
Some work in this direction has already been done \cite{Ventura:2009zz}, motivated
by ``hidden valley'' models \cite{Strassler:2006im}.

Important additional information is obtained if
the displaced vertex can be timed,
thus determining the velocity of $Y$.
Timing is probably more challenging and
depends on the final state and detector in question.
For decays involving muons, muon arrival times 
can determine the time of flight of $Y$.
This technique could also be used for other decay products
if the decay occurs in the muon system.
Timing in the hadronic or EM calorimeters can potentially extend the
kinds of vertices that can be timed.
Overcoming the experimental difficulty in timing measurement has a
significant payoff in event reconstruction, as we will see below.                  
 
%

Since the decay lengths are typically larger than a detector
size, we focus on the case where only one of the NLSPs
decays inside the detector.
We assume that the velocity of the NLSP that decays in the
detector has been determined as discussed above.
%
%
We need to determine the 4-momenta of the two $Y$ particles
to reconstruct the prompt event.
We assume that the visible particles and missing $p_T$
in the prompt event can be reconstructed separately from those
of the displaced vertex.
This may be possible only in favorable events where the hard
visible particles and missing $p_T$ from the prompt event 
do not point to activity associated with the displaced vertex.
The missing $p_T$ and the velocity of the NLSP
gives 5 constraints on the 8 components of the $Y$
4-momenta, so we still need
3 constraints for each such event.
If we have $n$ such events, we get
$2(2n - 1)$ constraints by imposing equality of the
masses of the ``mother'' particles $M$ (see Fig.~1b)
and the NLSPs $Y$.
For 2 such events, the prompt part of the event can be
completely reconstructed (up to discrete combinatoric ambiguities). 
If the prompt decays involve additional intermediate particles
we can get additional constraints \cite{Cheng:2008hk}
at the price of additional model-dependence.
%

More generally, the point is that the additional kinematic
information from displaced vertices is
very helpful in reconstructing the prompt part of the event.
As another example, a single event with two
displaced decays positioned allows full reconstruction
of the prompt event.


{\bf X reconstruction:}
We now turn to the reconstruction of $X$ properties.  The coupling between $X$ and $Y$ is in principle measurable via $Y$'s lifetime, indirectly given by 
the fraction of events with a decay in the detector,
but this parameter is not usually important for calculating the dark matter density.
On the other hand, the $X$ mass is crucial in determining whether 
the $X$ particle observed in a displaced vertex is
the dark matter.   
This requires the reconstruction of the total 4-momentum of the
visible decay products of the NLSP, $p_{\rm vis}$, which may be
highly nontrivial.

We first consider the case that the NLSP 4-momenta $p_{Y}$ is known,
either by measurement for a charged NLSP or determined by
kinematic reconstruction as discussed above.
For each such event, the invariant mass of
$p_{\rm inv} = p_{Y} - p_{\rm vis}$ then gives the mass of the
invisible decay products.
If there is only one invisible $X$ particle produced in each decay,
this will be the mass of $X$;
if there are $n$ $X$ particles plus possible neutrinos,
the mass distribution will have a lower endpoint at $n\, m_X$.
The shape of the distribution then gives a handle
on the value of $n$.  

If the NLSP 4-momenta is unknown, the situation is more difficult.
However, if its velocity can be determined by vertexing and timing,
we can determine $m_X$ if a single $X$
accounts for all of the missing energy.
Given the NLSP velocity, it is possible to boost to its center
of mass frame.
In this frame, by energy-momentum conservation we have 
\beq
m_{Y} = E_{\rm vis}+\sqrt{m_X^2+|\vec{p}_{\rm vis}|^2}
\eeq
where $E_{\rm vis}$, $\vec{p}_{\rm vis}$ are the energy and
momentum of the visible decay products in the NLSP rest frame.
Since this is an equation with two unknowns ($m_{Y}$, $m_X$),
two events may be sufficient to solve it.
Events involving decays to the same single
visible particle give degenerate information, and we only
get a single constraint.
However, if there are events with more than one kind of
visible particle ({\it e.g.\/}~Higgs and $Z$ production) the masses can be solved.
We conclude that by combining a small number of events
it is possible to solve for both the
NLSP and LSP masses without relying on kinematic reconstruction
of the prompt event.


\section{Conclusions}
In this Letter, we have discussed supersymmetric dark matter models
where the relic abundance is not due to standard WIMP cosmology,
specifically superWIMP and asymmetric dark matter models.
In these models, the dark matter is produced at colliders only
as a result of the decay of a NLSP with a macroscopic decay
length, and there is no guarantee that the properties of the 
dark matter can be probed at collider experiments.
However, for NLSP decay lengths not too far from the minimum
value of order 10~m, a significant number of NLSP decays can occur
inside the detector, giving rise to a rich set of possible
displaced vertices at colliders.

We have shown that measuring the position and/or time delay of the
displaced vertices is extremely useful for reconstructing the
SUSY mass spectrum,
while fully reconstructing the visible decay products of vertices can be
used to determine the dark matter mass.
For both superWIMP and asymmetric dark matter models
the mass is the crucial parameter that connects
cosmological observations of dark matter to collider data.
Quantitative studies of several scenarios are under investigation.
Of course the mass measurement techniques discussed here apply 
also to displaced decays to invisible particles that are not dark matter,
{\it e.g.\/}~light gravitinos \cite{Reece}
or neutrinos \cite{Chang:2009dh}.
This strongly motivates overcoming the experimental challenges
in the study of highly displaced vertices at colliders.


\begin{acknowledgments}
This work was supported in part by DOE grant
DE-FG02-91-ER40674.
We thank the organizers and participants of the University of Washington
workshop on long lived exotic particles
(supported by DOE grant DE-FG02-96-ER40956),
for motivation and feedback on this work.
\end{acknowledgments}

\bibliography{displacedDM.bib}

\end{document}